\documentclass[journal,a4paper,onecolumn]{IEEEtran}

\usepackage{amsfonts,amsmath,amssymb,mathrsfs,amsbsy,amsthm}
\usepackage{tikz}
\usepackage{subfigure}
\usepackage{color}
\usepackage{graphicx,epsfig}
\usepackage{stfloats} 
\usepackage{hyperref}
\usepackage{psfrag}

\newcounter{example}
\newenvironment{example}[1][]{\refstepcounter{example}\par\medskip\noindent%
   \textbf{Example~\theexample. #1} \rmfamily}{\medskip}
\newcounter{examplec}
\newenvironment{examplec}[1][]{\refstepcounter{examplec}\par\medskip\noindent%
   \textbf{Example~\theexamplec} {\em Continued. #1} \rmfamily}{\medskip}

\newcounter{algorithm}

\newcommand{\cX}{{\cal X}}

\newtheorem{thm}{Theorem}

\begin{document}

\sloppy

\title{
An Improved Sub-Packetization Bound for \\ Minimum Storage Regenerating Codes\thanks{The material in this paper will be presented in part at the IEEE International Symposium on Information Theory (ISIT 2013), Istanbul, Turkey, July 2013.}
}



\author{\IEEEauthorblockN{
Sreechakra Goparaju\IEEEauthorrefmark{1},
Itzhak Tamo\IEEEauthorrefmark{2},
and
Robert Calderbank\IEEEauthorrefmark{4}}

\IEEEauthorblockA{\IEEEauthorrefmark{1}Department of Electrical Engineering,
Princeton University, USA}

\IEEEauthorblockA{\IEEEauthorrefmark{2}Department of Electrical and Computer Engineering, University of Maryland, USA} 

\IEEEauthorblockA{\IEEEauthorrefmark{4}Department of Electrical Engineering, Duke University, USA}}




\newtheoremstyle{mystyle}
  {}
  {}
  {\itshape}
  {}
  {\bfseries}
  {\,}
  { }
  {}

\theoremstyle{mystyle}

\maketitle

\begin{abstract}
Distributed storage systems employ codes to provide resilience to failure of multiple storage disks.
Specifically, an $(n, k)$ MDS code stores $k$ symbols in $n$ disks such that the overall system is tolerant to a failure of up to $n-k$ disks. However, access to at least $k$ disks is still required to repair a single erasure. To reduce repair bandwidth, array codes are used where the stored symbols or packets are vectors of length $\ell$. 
MDS array codes have the potential to repair a single erasure using a fraction $1/(n-k)$ of data stored in the remaining disks.
We introduce new methods of analysis which capitalize on the translation of the storage system problem into a geometric problem on a set of operators and subspaces.
In particular, we ask the following question: for a given $(n, k)$, what is the minimum vector-length or sub-packetization factor $\ell$ required to achieve this optimal fraction? 
For \emph{exact recovery} of systematic disks in an MDS code of low redundancy, i.e. $k/n > 1/2$, the best known explicit codes \cite{WTB12} have a sub-packetization factor $\ell$ which is exponential in $k$. It has been conjectured \cite{TWB12} that for a fixed number of parity nodes, it is in fact necessary for $\ell$ to be exponential in $k$. In this paper, we provide a new log-squared converse bound on $k$ for a given $\ell$, 
and prove that $k \le 2\log_2\ell\left(\log_{\delta}\ell+1\right)$, for an arbitrary number of parity nodes $r = n-k$, where $\delta = r/(r-1)$. 
\end{abstract}

\section{Introduction}\label{sec:intro}
Maximum Distance Separable (MDS) codes are ubiquitous in distributed storage systems \cite{DRWS11} since they provide the maximum resilience to erasures for a given redundancy.
We define an $(n,k,\ell)$ storage system as consisting of $n$ nodes or disks of capacity $\ell$ (data) units, and storing a total of $k\ell$ data units. When the $\ell$ units in each disk constitute a symbol in an MDS (array) code, the system is immune to an erasure of up to $r=n-k$ disks. Failure of a single disk at a time occurs most frequently in practice. The objective is to quickly and efficiently recover the data in an erased disk. A na{\"i}ve way is to reconstruct the entire data by using any $k$ of the surviving disks and recover the data in the lost node. However, as can be seen in Example \ref{ex:simple}, 
transmission of $k\ell = 2 \times 2 = 4$ units to the repair center is not necessary to recover the loss of $\ell = 2$ data units; transmission of $3$ units is sufficient. Dimakis et al. \cite{DGWWR10} formalized this problem of efficient repair and proved that the bandwidth or the amount of transmitted data required to recover a single disk erasure in an MDS code is lower bounded by
\begin{eqnarray*}
\left(\frac{n-1}{n-k}\right)\ell &=& \left(\frac{n-1}{r}\right)\ell \textrm{   data units},
\end{eqnarray*}
where all the surviving $n-1$ disks transmit a fraction $1/r$ of their data.

\begin{figure}
\centering
\begin{tikzpicture}[scale=0.4,>=stealth]

\draw (0,0) circle (1cm); \draw (0,0) node {$v_1$}; \draw[red] (1,1) -- (-1,-1); \draw[red] (1,-1) -- (-1,1);
\draw (0,-3.5) circle (1cm); \draw (0,-3.5) node {$v_2$};
\draw[blue] (0,-7) circle (1cm); \draw (0,-7) node {$v_3$};
\draw[blue] (0,-10.5) circle (1cm); \draw (0,-10.5) node {$v_4$};

\draw[gray,->] (1,-3.5) .. controls (3,-3) and (3,-2) .. (1,0); \draw[below] (2.5,-3.3) node {$S_{1,2}v_2$};
\draw[gray,->] (1,-7) .. controls (5,-6) and (5,-3) .. (1,0); \draw[below] (2.5,-6.8) node {$S_{1,3}v_3$};
\draw[gray,->] (1,-10.5) .. controls (7,-9) and (7,-3) .. (1,0); \draw[below] (2.5,-10.3) node {$S_{1,4}v_4$};

\draw[blue] (-14,-11.5) rectangle (-2,-9.5); \draw (-11,-10.5) node {$a_2 + b_1$}; \draw (-5,-10.5) node {$a_1 + a_2 + b_2$}; \draw[blue] (-8,-11.5) -- (-8,-9.5);
\draw[blue] (-14,-8) rectangle (-2,-6); \draw (-11,-7) node {$a_1 + b_1$}; \draw (-5,-7) node {$a_2 + b_2$}; \draw[blue] (-8,-8) -- (-8,-6);
\draw (-14,-4.5) rectangle (-2,-2.5); \draw (-11,-3.5) node {$b_1$}; \draw (-5,-3.5) node {$b_2$}; \draw (-8,-4.5) -- (-8,-2.5);
\draw (-14,-1) rectangle (-2,1); \draw (-11,0) node {$a_1$}; \draw (-5,0) node {$a_2$}; \draw (-8,-1) -- (-8,1);
\end{tikzpicture}
\caption{A $(4,2,2)$ MDS code over $\mathbb{F}_2$ (\cite{DRWS11,BBBM94}), where nodes $1$ and $2$ store the systematic data (in black) and nodes $3$ and $4$ store the parity data (in blue). For example, node $2$ stores the information vector $v_2=(b_1,b_2)^t$. Any single erasure can be recovered using $3$ data units. The figure shows the repair scenario of node $1$. Each of the remaining nodes $v_j, j \neq 1$, transmits to the repair center $S_{1,j}v_j$ to aid in the repair process. For example, the nodes can transmit their second data units $b_2$, $a_2+b_2$ and $a_1+a_2+b_2$ to recover $v_1 = (a_1, a_2)$.
To recover the second parity node $(a_2+b_1,a_1+a_2+b_2)$, we can use $a_1, b_1+b_2$ and $a_2+b_2$, where the second unit of information can be obtained from the second systematic node by a linear combination of its stored data.}
\label{fig:mds}
\end{figure}
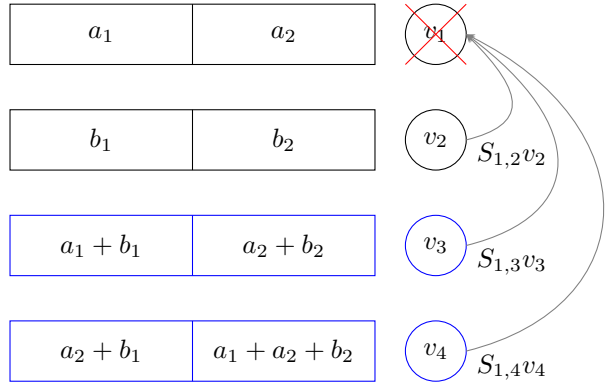


Codes which achieve this lower bound are called \emph{optimal bandwidth MDS codes} or \emph{minimum-storage regenerating (MSR) codes} with $n-1$ \emph{helper} nodes. Much progress has been made recently in constructing such codes. Network coding is sufficient to obtain optimal bandwidth codes for \emph{functional} repair, where the objective is to recover a lost disk such that the MDS property is preserved in the new set of $n$ disks. The case of \emph{exact} repair, requiring the recovered node to exactly replicate the lost disk has proved to be more challenging. Optimal bandwidth exact repair codes were constructed in \cite{SR10,RSK11} for the low-rate regime ($k/n \le 1/2$), where the number of parity nodes exceeds the number of systematic nodes. In the high-rate case ($k/n > 1/2$), optimality was proved to be achievable asymptotically \cite{CJMRS13}. Recent contributions construct finite length (finite $\ell$) codes which exactly recover the systematic nodes \cite{PDC11, CHLM11, WTB12}. Explicit finite-length optimal exact repair codes have also been constructed \cite{PDC11, WTB11},
where both systematic as well as parity nodes are recoverable optimally.

Before we can apply our methods we need to translate from the language of computer storage to the language of operators and subspaces\footnote{Note that this paper covers only linear MDS codes and linear repair.}, and we do this via an example.
\begin{example}[$(4,2,2)$ MDS code over $\mathbb{F}_2.$]\label{ex:simple}
Refer to Fig. \ref{fig:mds}.
The first two nodes store the systematic data units $v_1$ and $v_2$, each a binary column-vector of length two.
The last two nodes are the parity nodes, which can be represented in the following generic form:
\begin{eqnarray*}
\begin{array}{ccccc}
v_3 & = & A_{1,1}v_1 + A_{1,2}v_2 & = & \left(\begin{array}{cc} 1&0\\0&1 \end{array}\right) v_1 + \left(\begin{array}{cc} 1&0\\0&1 \end{array}\right) v_2,
\\[3mm]
v_4 & = & A_{2,1}v_1 + A_{2,2}v_2 & = & \left(\begin{array}{cc} 0&1\\1&1 \end{array}\right) v_1 + \left(\begin{array}{cc} 1&0\\0&1 \end{array}\right) v_2,
\end{array}
\end{eqnarray*}
where the matrices or linear operators $A_{i,j}, i, j \in \{1,2\}$, determine how the systematic data is encoded in the parity nodes so that the overall code is MDS in nature. Suppose node $1$ fails and a replacement node contacts the remain three nodes to restore $v_1$.  From Fig. \ref{fig:mds}, we know that one way to do so is for the remaining nodes to send their second bit. In general, however, suppose each node sends a bit which is a linear combination of the data stored in it. If node $j$ sends the bit $S_{i,j}v_j$ to aid in the recovery of $v_i$, $S_{i,j} \in \mathbb{F}_2^{1 \times 2}$, we have the following ``information'' at the replacement node $1$: 
\begin{eqnarray*}
\left(
\begin{array}{c}
S_{1,2}v_2\\
S_{1,3}A_{1,1}v_1+S_{1,3}A_{1,2}v_2\\
S_{1,4}A_{2,1}v_1+S_{1,4}A_{2,2}v_2
\end{array}
\right) &=&
\left(\begin{array}{cc}0&S_{1,2}\\
S_{1,3}A_{1,1}&S_{1,3}A_{1,2}\\
S_{1,4}A_{2,1}&S_{1,4}A_{2,2}\end{array}\right)
\left(\begin{array}{c}v_1\\v_2 \end{array}\right).
\end{eqnarray*}

To recover $v_1$ optimally from the available information, therefore, it is necessary and sufficient that the matrices
\begin{eqnarray*}
\left(\begin{array}{c}S_{1,3}A_{1,1}\\S_{1,4}A_{2,1}\end{array}\right)\textrm{ and }\left(\begin{array}{c}S_{1,2}\\S_{1,3}A_{1,2}\\S_{1,4}A_{2,2}\end{array}\right)
\end{eqnarray*}
are of rank $2$ and $1$, respectively. 
It is easy to see that another way of viewing the necessary and sufficient conditions is that $S_{1,2}, S_{1,3}$ and $S_{1,4}$ are one-dimensional subspaces such that $S_{1,2}, S_{1,3}A_{1,2}$ and $S_{1,4}A_{2,2}$ are the same subspace, and $S_{1,3}A_{1,1}$ is a complementary subspace to $S_{1,4}A_{2,1}$. Notice that here, we use the same notation $S_{i,j}$ to represent both the $1 \times 2$ matrix and the one-dimensional subspace corresponding to the span of its row vector. In the example in Fig. \ref{fig:mds}, the subspaces $S_{1,2}, S_{1,3}$ and $S_{1,4}$ are all given by the same matrix, $\left(0, 1\right)$. 
The subspaces $S_{i,j}$ will be appropriately referred to as the {\em repairing subspaces}. 
\hfill $\square$
\end{example}

\subsection{Bounds on Sub-Packetization:}
In this paper, we look closely at the relationship between the disk capacity $\ell$ and the number of systematic nodes for an optimal bandwidth MDS code for a given number of parity nodes $r$.
This question is intimately connected to the concepts of array and block codes \cite{BBBM94, BBV96}, linear vector coding, sub-packetization \cite{CJMRS13}, and symbol extension \cite{CJ08}. The capacity $\ell$, also known as the \emph{sub-packetization factor}, represents the minimum dimension over which all the recovery arithmetic operations are needed, independent of the field $\mathbb{F}$ involved. For example, if $\ell=1$, namely, each node stores exactly one symbol, then we cannot do better than reconstructing the entire file to recover a single erasure (because an ability to do better would violate the MDS property).
Hence, a sub-packetization of $\ell > 1$ is required to achieve the optimal bandwidth property. 

The disk capacity of $\ell$ symbols is linked to its actual ``raw capacity'' (or size in bits or bytes) via the field $\mathbb{F}$ over which the MDS code and the repair subspaces are assumed. For a given disk capacity of $\ell$ symbols and a given number of parity nodes $r$, the number of systematic nodes $k$ which a DSS can ``support'' while storing an optimal bandwidth MDS code is bounded from above. Alternatively, a disk of size $\ell$ bits cannot store more than $\ell$ symbols (in any field) and therefore has a field-independent upper bound on $k$. 

\begin{example}[A $1$kB Storage Disk.] Consider a storage disk of size $1$ kilobyte $=$ $2^{13}$ bits. This disk can have a capacity of at most $\ell = 2^{13}$ symbols. For two parity nodes, the current upper bound on $k$ is exponential in $\ell$ and amounts to more than $10^{2467}$ systematic nodes. We prove that $k$ in fact cannot be more than $365$.
\end{example}

For the low-rate case, i.e, when $k/n \le 1/2$, a linear sub-packetization (in terms of $r$) is sufficient \cite{RSK11}. In fact, $\ell = n-k = r$ when all the $n-1$ nodes aid in repair, because each disk need only contribute one unit (scalar repair) of repair bandwidth. The absence of optimal scalar linear repair codes \cite{SRKR10} for the case of $k/n > 1/2$ justifies the search for vector linear repair codes. Cadambe et al. incorporated the idea of symbol-extension from interference alignment \cite{CJMRS13} and proved the existence of exact-repair MSR codes, albeit for asymptotically large $\ell$ for a fixed number of parity nodes $r$. Finite-length codes were discovered by \cite{PDC11, CHLM11, WTB12} where $\ell$ is exponential in the number of systematic nodes $k$. Can one do better (lower) than an exponential $\ell$? Conversely, can one have more than a logarithmic number of systematic nodes $k$ for a given disk capacity $\ell$ and $r$?

The example in Fig. \ref{fig:mds} is an MDS code with $k = 2$ systematic nodes for $\ell = 2$ and $r = 2$. Table \ref{table:mds} shows an MDS code \cite{TWB12} with $k = 4$ systematic nodes for the same $\ell$ and $r$ over the field $\mathbb{F}_7$.
It might be expected that if we increase the field size, we can arbitrarily increase $k$. But as shown in \cite{TWB12}, $k$ cannot be more than $5$ even if the field size is infinite.

\begin{table}[ht]
\centering
\normalsize
\begin{tabular}{|c||c|c|}
\hline
 $v_1$ & $a$ & $w$\\
 \hline
 $v_2$ & $b$ & $x$\\
 \hline
 $v_3$ & $c$ & $y$\\
 \hline
$v_4$ & $d$ & $z$\\
\hline
$v_5$ & $a+b+c+d$ & $w+x+y+z$ \\
\hline
$v_6$ & $a+5w+b+2c+5d$ & $3w+2b+3x+4y+5z$\\
\hline
\end{tabular}
\caption{A $(6,4,2)$ MDS code over $\mathbb{F}_7$}
\label{table:mds}
\end{table}

%

Tamo et al. \cite{TWB12} conjectured that for a given disk capacity $\ell$ and an arbitrary fixed number of parity nodes $r$, the maximum number of systematic nodes $k$ is of the order of $\log\ell$. 
In the sequel, we provide a significant improvement to the bounds \cite{TWB13} on the largest number of systematic nodes $k$ for given values of $\ell$ and $r$:
\begin{eqnarray*}
(r+1)\log_r\ell \le k \le \ell{\ell \choose \ell/r}.
\end{eqnarray*}

\subsection{Our Contribution:}
The existence of optimal bandwidth MDS codes for the exact repair of systematic nodes can be expressed as an interesting linear algebra problem (e.g. \cite{TWB12}) involving certain subspaces and linear operators, as described in Section \ref{sec:problem}. This problem is a simplified version of all such rank conditions as mentioned in Example \ref{ex:simple}. We introduce new methods of analysis which capitalize on this restricted geometry of operators and subspaces.
In particular, we use these methods to provide new upper bounds on the maximum number of systematic nodes possible $k$ for a given sub-packetization factor $\ell$. 

We introduce ideas progressively in subsequent sections which will help us gradually narrow the gap between the upper and lower bounds on $k$ and also hopefully make the proofs more comprehensible and intuitive. Wherever possible, we also restrict the proofs to the case of two parity nodes ($r = 2$) and provide the proof for an arbitrary number of parity nodes in the appendix.
Exploiting the subspace conditions obtained in Section \ref{sec:problem}, we first prove in Section \ref{sec:ub1} that for an arbitrary number of parity nodes $r$, the maximum number of systematic nodes $k$ is bounded by
\begin{eqnarray}
k &\le& \ell^2,
\end{eqnarray}
or in other words, the sub-packetization factor should at least be $\sqrt{k}$. For the special case of $2$ parity nodes, we then derive a stronger upper bound on $k$ in Section \ref{sec:ub2}:
\begin{eqnarray}
k &\le& 4\ell +1,
\end{eqnarray}
i.e., the sub-packetization is required to be of the order of $k$. 
Finally, in Section \ref{sec:ub3}, we prove that for $r$ parity nodes, we can extend the ideas in the previous sections to strengthen the bound to:
\begin{eqnarray}
k &\le& 2\left(\log_2\ell\right)\left(\log_{\delta}\ell+1\right)+1,
\end{eqnarray}
where $\delta = r/(r-1)$.

We note here that the results on the geometry of operators and subspaces derived in this paper are of independent interest. For instance, under circumstances similar to the problem in this paper, that is, for an optimal bandwidth MDS code for the exact repair of systematic nodes, optimal secure data rates can be derived. The maximum file size which can be securely stored on the DSS has been derived in \cite{GRCP13}, when an eavesdropper has access to the repair data for any given number of systematic nodes.


\section{Problem Setting}\label{sec:problem}
We define an $(n,k,\ell)$ MDS array code as a set of $n$ symbol vectors (disks) of length $\ell$ over a field $\mathbb{F}$, such that any set of $k$ vectors are sufficient to recover the entire data of $k\ell$ units. The first $k$ symbols represent the systematic nodes consisting of the data vectors $v_1, \ldots, v_k$ of column-length $\ell$. Each of the remaining $r = n-k$ symbols is a parity node, which stores a linear combinations of the systematic data vectors. More formally, the vector $v_{k+i}$ stored in parity node $i$ is given by
\begin{eqnarray*}
v_{k+i} = \sum_{j=1}^kA_{i,j}v_j,
\end{eqnarray*}
where $A_{i,j}$ is a square \emph{encoding matrix} of order $\ell$ 
corresponding to the parity node $i \in [r] := \{1,2,\ldots,r\}$ and the systematic node $j \in [k]$. For optimal (bandwidth) repair of a failed systematic node $i \in [k]$, all other nodes transmit a fraction $1/r$ of the stored data, i.e., the \emph{helper} node $j \neq i, j \in [n]$ transmits a vector of length $\ell/r$ given by $S_{i,j}v_{j}$, where $S_{i,j}$ is a matrix in $\mathbb{F}^{\ell/r \times \ell}$. An alternate interpretation\footnote{We alternate between interpreting $S_{i,j}$ and $S_{i,k+t}A_{t,j}$ as matrices of size $\ell/r \times \ell$ and subspaces of their row spans of dimension $\ell/r$.} is that the vector transmitted by node $j$ is the projection of $v_j$ onto a subspace of dimension $\ell/r$. The subspace corresponds to the subspace spanned by the rows of $S_{i,j}$. 
It can be shown using interference alignment ideas \cite{WTB12} that the optimal repair of a systematic node $i$ is possible if and only if there exist $\ell/r \times \ell$ matrices $S_{i,j}, j \neq i, j \in [n]$, which satisfy the following \emph{subspace properties}:
\begin{eqnarray}\label{sp1}
S_{i,j} &\backsimeq& S_{i,k+t}A_{t,j},\mathrm{\,\,\,and}\\
\label{sp2}\sum_{t=1}^rS_{i,k+t}A_{t,i} &\backsimeq& \mathbb{F}^{\ell},
\end{eqnarray}
for all $j \neq i, j \in [k], t \in [r]$. The equalities $\backsimeq$ in the subspace properties are defined on the row spans (or subspaces) instead of the corresponding matrices. The sum of subspaces $B$, $C$ is defined as $B+C = \{b+c:b\in B, c \in C\}$. Since the dimension of each subspace $S_{i,j}$ is $\ell/r$, it is clear from (\ref{sp2}) that each encoding matrix $A_{t,i}$ is invertible and that the sum of the subspaces is a direct sum.

\begin{examplec}
As an illustrative example,
we derive the repairing subspaces for the $(4,2,2)$ MDS code over the binary field as described in Example \ref{ex:simple} and Fig. \ref{fig:mds}. For the given encoding matrices, we know that the subspaces $S_{1,2}, S_{1,3}$ and $S_{1,4}$ are the same subspace, say, $S_1$. We also know that $S_{1,3}$ is a complementary subspace to $S_{1,4}A_{2,1}$, that is, $S_1$ is complementary to $S_1A_{2,1}$. By a simple calculation, we find that $S_1$ can be any of the three possible one-dimensional subspaces -- $\mathrm{span}\{(0,1)\}, \mathrm{span}\{(1,0)\}$, and $\mathrm{span}\{(1,1)\}$. Fig. \ref{fig:mds} uses the first of these.

If the second systematic node storing $v_2$ is to be recovered, the repairing subspaces are $S_{2,1}, S_{2,3}$ and $S_{2,4}$. The subspaces $S_{2,1}, S_{2,3}$ and $S_{2,4}A_{2,1}$ represent the same subspace, whereas $S_{2,3}$ and $S_{2,4}$ must be complementary to each other. Again, it can easily be checked that there are three possible scenarios, where $S_{2,1}$ can be represented by any of the three non-zero $1 \times 2$ binary matrices, and $S_{2,3} = S_{2,1}$ and $S_{2,4} = S_{2,1}A_{2,1}^{-1}$.
\hfill $\square$
\end{examplec}


The following theorem proved in \cite{TWB13} significantly simplifies the algebraic problem consisting of the subspace properties (\ref{sp1}) and (\ref{sp2}) by transforming it into one for which the repairing subspaces do not depend on the helper node.

\begin{thm}[{{\cite[Theorem 2]{TWB13}}}]
\label{thm-constant-repairing-subspaces}
If there exists an optimal bandwidth $(k+r+1, k+1,\ell)$ MDS code, then there exists an optimal bandwidth $(k+r, k,\ell)$ MDS code where the repairing subspaces 
are independent of the helper node. In other words, there exist subspaces $S_1, \ldots, S_k$, and encoding matrices $A_{t,j}, t \in [r], j \in [k]$, which satisfy the following subspace properties:
\begin{eqnarray}
\label{sptx1}
S_i &\backsimeq& S_iA_{t,j}, \,\, \textrm{ and}\\
\label{sptx2}
\sum_{u=1}^{r}S_iA_{u,i} &\backsimeq& \mathbb{F}^{\ell},
\end{eqnarray}
for all distinct $i, j \in [k]$.
\end{thm}


Furthermore, it can be shown \cite{TWB12,TWB13} that the each of the encoding matrix of one of the parity nodes in the transformed MDS code can be assumed to be the identity matrix. 
In particular, 
if there exists an optimal bandwidth $(k+3,k+1,\ell)$ MDS code, then there exist a set of invertible matrices $\Phi_1, \ldots, \Phi_k$ of order $\ell$ and a corresponding set of subspaces $S_1, \ldots, S_k$, each of dimension $\ell/2$ such that for any distinct $i, j \in [k]$,
\begin{eqnarray}
\label{subspaceconditions1}
S_i\Phi_j &\backsimeq& S_i, \mathrm{\,\,\,and}\\
\label{subspaceconditions2}
S_i\Phi_i + S_i &\backsimeq& \mathbb{F}^{\ell},
\end{eqnarray} 
where the sum can again be seen as a direct sum. 

In general, we want to find for a given sub-packetization factor $\ell$ and a given number of parity nodes $r$, the largest number of systematic nodes $k$ for which there exists an optimal repair scheme (of systematic nodes). 


\section{Primer: A Simple Sub-Packetization Bound}
\label{sec:ub1}
We first prove a quadratic (in terms of $\ell$) upper bound for the maximum possible number of systematic nodes $k$ for the general case of $r$ parity nodes.
\begin{thm}\label{thm1} 
For a given disk capacity $\ell$ and for an arbitrary 
number of parity nodes $r$, the number of systematic nodes $k$ is upper bounded by
\begin{eqnarray}
k \le \ell^2.
\end{eqnarray}
\end{thm}
\begin{IEEEproof}
Consider the subspace properties (\ref{sp1}) and (\ref{sp2}) for the parity nodes $1$ and $2$. The idea\footnote{
The idea of making such a translation appears in \cite{TWB12}, again in {\cite[Theorem 2]{TWB13}}, and is included here to make the proof self-contained. 
It can also be shown from the MDS property of the system that all the coding matrices $A_{i,j}$ have full rank, and from the subspace conditions that all the subspaces $S_{i.j}$ have full rank as well.
} is to convert the given set of subspaces and matrices to another set which satisfy properties similar to the second set of subspace properties (\ref{subspaceconditions1}) and (\ref{subspaceconditions2}). Let us define for $i \in [k-1]$,
\begin{eqnarray*}
\Theta_i = A_{1,i}A_{2,i}^{-1}A_{2,k}A_{1,k}^{-1}.
\end{eqnarray*}
We then have, using the properties (\ref{sp1}) and (\ref{sp2}), for $j \not \in \{i,k\}$,
\begin{eqnarray}
S_{j,k+1}\Theta_i &\backsimeq& S_{j,k+1}A_{1,i}A_{2,i}^{-1}A_{2,k}A_{1,k}^{-1}\nonumber\\
&\backsimeq& S_{j,k+2}A_{2,i}A_{2,i}^{-1}A_{2,k}A_{1,k}^{-1} \nonumber\\
&\backsimeq& S_{j,k+2}A_{2,k}A_{1,k}^{-1} \nonumber\\
&\backsimeq& S_{j,k+1}A_{1,k}A_{1,k}^{-1} \nonumber\\
&\backsimeq& S_{j,k+1}, \label{spp1}
\end{eqnarray}
and
\begin{eqnarray}
&& S_{i,k+1}\Theta_i \cap S_{i,k+1} = \{0\} \label{spp2}\\
 &\iff& S_{i,k+1}A_{1,i}A_{2,i}^{-1} \cap S_{i,k+1}A_{1,k}A_{2,k}^{-1} = \{0\}\nonumber\\
 &\iff& S_{i,k+1}A_{1,i}A_{2,i}^{-1} \cap S_{i,k+2}A_{2,k}A_{2,k}^{-1} = \{0\}\nonumber\\
 &\iff& S_{i,k+1}A_{1,i} \cap S_{i,k+2}A_{2,i}= \{0\}\nonumber,
\end{eqnarray}
which is indeed true by (\ref{sp2}). Defining $S_{i,k+1}$ as $S_i$, we have the properties
\begin{eqnarray}
\label{sc1}
S_i\Theta_j &\backsimeq& S_i, \mathrm{\,\,\,and}\\
\label{sc2}
S_i\Theta_i \cap S_i &=& \{0\},
\end{eqnarray}
for any distinct $i, j \in [k-1]$, and $0$ is the zero-vector.

The proof follows from the observation that the matrices $\Theta_i, i \in [k-1]$ and the identity matrix $\mathrm{I}$ are linearly independent. Since they lie in the $\ell^2$-dimensional space of matrices $\mathbb{F}^{\ell \times \ell}$, we have $k \le \ell^2$. To see that the matrices are linearly independent, suppose they are not. Then without loss of generality we have some equation of the form,
\begin{eqnarray*}
\Theta_1 = \sum_{i=2}^t\alpha_i\Theta_i + \beta\mathrm{I}, \,\, t \le k, \,\alpha_i \neq 0.
\end{eqnarray*}
Then, operating $\Theta_1$ on $S_1$, we have
\begin{eqnarray*}
S_1\Theta_1 &\backsimeq& S_1\left(\sum_{i=2}^t\alpha_i\Theta_i + \beta\mathrm{I}\right).
\end{eqnarray*}
The right hand side of the above equation lies in $S_1$ from (\ref{sc1}) and the fact that any subspace is invariant under the identity transformation\footnote{A subspace $S$ is invariant under a linear operator $\Theta$ if $S\Theta \subseteq S$. If $S$ is invariant under two matrices, it is also invariant under their sum. If two subspaces are invariant under an operator, so is their intersection.}. But $S_1$ is non-intersecting\footnote{We use the word \emph{non-intersecting} in the context of subspaces to imply that they intersect only in the zero vector.} with the subspace on the left hand side $S_1\Theta_1$ from (\ref{sc2}), which implies that $S_1\Theta_1 = \{0\}$. This further implies that $S_1 = \{0\}$, by the non-singularity of $\Theta_1$, a contradiction to the fact that all $S_{i,k+1}$ are full rank matrices of rank $\ell/r$.
\end{IEEEproof}

\section{An Improved Bound for Two Parity Nodes}\label{sec:ub2}
In this section we prove a stronger upper bound on $k$ in the case of $r = 2$ parity nodes. By analyzing the geometry of the operators and the corresponding subspaces in two complementary directions, we prove that the number of systematic nodes $k$ can be upper bounded either by $4\ell$ or $8\log_2\ell$. 
We first investigate how the geometry of subspace intersection is related to the linear independence of corresponding operators.


Let $S_i \text{ and }\Phi_i, i\in [2t]$ be subspaces and matrices that satisfy \eqref{subspaceconditions1} and\eqref{subspaceconditions2}. Define $T$ to be the set of products of pairs of matrices $\Phi_i\Phi_j$, where $i$ is odd and $j$ is even. Clearly the cardinality of $T$ is $t^2.$ The following theorem shows a connection between linear dependencies of the matrices in $T$ and the subspaces $S_i$.

\smallskip
\newtheorem{lem}{Lemma}
\newtheorem{cor}{Corollary}
\begin{thm}\label{thm-lincomb2subsp}
If there exists an element in $T$, say $\Phi_i\Phi_j$, which can be expressed as a non-zero linear combination of other elements in $T$, then $S_i \cap S_j = \{0\}$. In other words, if $\Phi_i\Phi_j$ lies in the span of the rest of the elements in $T$, then $S_i$ and $S_j$ are complementary subspaces of dimension $\ell/2$. 
\end{thm}
\begin{IEEEproof}
Without loss of generality, let $(i,j)=(1,2)$. If $\Phi_1\Phi_2$ is in the span of the rest of the elements in $T$, then
\begin{eqnarray*}
\Phi_1\Phi_2 &=& \sum_{j \neq 2}\alpha_{1j}\Phi_1\Phi_j + \sum_{i \neq 1}\alpha_{i2}\Phi_i\Phi_2 + \sum_{\substack{m \neq 1\\n \neq 2}}\alpha_{mn}\Phi_m\Phi_n,
\end{eqnarray*}
where no $\alpha \in \mathbb{F}$ is zero in the above summation. Applying both sides to the subspace $S_2$, we obtain
\begin{eqnarray*}
S_2\left(\Phi_1\Phi_2 - \sum_{i \neq 1}\alpha_{i2}\Phi_i\Phi_2\right)\hspace{3cm}\\
 \backsimeq S_2\left(\sum_{j \neq 2}\alpha_{1j}\Phi_1\Phi_j + \sum_{\substack{m \neq 1\\n \neq 2}}\alpha_{mn}\Phi_m\Phi_n\right).
\end{eqnarray*}
By properties (\ref{subspaceconditions1}), (\ref{subspaceconditions2}), the left hand side of the above equation lies in $S_2\Phi_2$, whereas the right hand side lies in $S_2$. This is possible only if
\begin{eqnarray*}
S_2\left(\Phi_1\Phi_2 - \sum_{i \neq 1}\alpha_{i2}\Phi_i\Phi_2\right) = \{0\}.
\end{eqnarray*}
Because $\Phi_2$ is full rank, this reduces to
\begin{eqnarray*}
S_2\left(\Phi_1 - \sum_{i \neq 1}\alpha_{i2}\Phi_i\right) = \{0\}.
\end{eqnarray*}
Thus, the relation also holds for a subspace in $S_2$:
\begin{eqnarray}\label{aux1}
\left(S_1\cap S_2\right)\left(\Phi_1 - \sum_{i \neq 1}\alpha_{i2}\Phi_i\right) &=& \{0\}, \,\mathrm{or}\\
\left(S_1\cap S_2\right)\left(\sum_{i \neq 1}\alpha_{i2}\Phi_i\right) &=& \left(S_1\cap S_2\right)\Phi_1.\nonumber
\end{eqnarray}
As before, by properties (\ref{subspaceconditions1}), (\ref{subspaceconditions2}), the right hand side of the above equation lies in $S_1\Phi_1\cap S_2$, whereas the left hand side lies in $S_1\cap S_2$. Note that on the left hand side, $i \neq 2$ by construction. 
Because $S_1\Phi_1 \cap S_1 = \{0\}$, so is $(S_1\Phi_1\cap S_2) \cap (S_1\cap S_2)$. Therefore, (\ref{aux1}) is possible only if 
\begin{eqnarray*}
\left(S_1 \cap S_2\right) \Phi_1 &=& \{0\},\,\mathrm{i.e.}\\
S_1 \cap S_2 &=& \{0\},
\end{eqnarray*}
because of the non-singularity of $\Phi_1$.
\end{IEEEproof}

\smallskip

\begin{cor}\label{cor1}
If for any operator $\Phi_i\Phi_j$ in $T$ the corresponding subspace $S_i \cap S_j$ does not equal to $\{0\}$, then $T$ consists of $t^2$ linearly independent elements.
\end{cor}


\smallskip
Corollary \ref{cor1} bounds the size of collections of operators for which the corresponding subspaces intersect non-trivially in pairs. This result can similarly be extended for non-trivial intersection of triples and higher tuples of subspaces. We do not use this generalization to obtain upper bounds on the number of systematic nodes $k$ in this paper, and therefore relegate it to Appendix \ref{appendix:cor1}. We believe that this generalization may be of independent interest.
The next result will bound the number of operator pairs for which the corresponding subspaces intersect trivially.
\smallskip

\begin{thm}\label{thm-completingsubsp2}
Suppose we have a set of $n \le k/2$ disjoint pairs of subspaces $(S_i, S_j)$, such that $S_i \cap S_j = \{0\}, \{i,j\} \in [k]$. Then we can find a set of $2^n$ linearly independent matrices in $\mathbb{F}^{\ell \times \ell}$.
\end{thm}
\begin{IEEEproof}
Without loss of generality, let these pairs be $(S_1,S_2), (S_3,S_4), \ldots, (S_{2n-1},S_{2n})$.
Let $M$ be the following set of $2^n$  $\ell \times \ell$ matrices: 
\begin{eqnarray*}
\Upsilon_{\epsilon_1\epsilon_2\ldots\epsilon_n}=\prod_{j=1}^{n}\left(\Phi_{2j-1}\Phi_{2j}\right)^{\epsilon_j},\,\,\,\mathrm{where}\,\,\epsilon_j \in \{0,1\} \,\,\forall j \in [n],
\end{eqnarray*}
where the product goes from left to right. For example, 
\begin{eqnarray*}
\Upsilon_{11} = \left(\Phi_1\Phi_2\right)\left(\Phi_3\Phi_4\right).
\end{eqnarray*}
We now prove by induction that no element in $M$ lies in the (linear) span of its remaining elements. The induction is on the sets $M_1, M_2, \ldots, M_{n-1}, M_n (= M)$, where $M_s$ is the set of the following $2^s$ $\ell \times \ell$ matrices:
\begin{eqnarray*}
\Upsilon_{\epsilon_1\epsilon_2\ldots\epsilon_s}=\prod_{j=1}^{s}\left(\Phi_{2j-1}\Phi_{2j}\right)^{\epsilon_j},\,\,\,\mathrm{where}\,\,\epsilon_j \in \{0,1\} \,\,\forall j \in [s].
\end{eqnarray*}

\smallskip
\emph{Induction Claim:} For all $s \in [n]$, no element in the set $M_s$ lies in the span of its remaining $2^{s}-1$ elements.

\emph{Base case:} For $s=1$, the proof follows from the fact that $\Upsilon_0 = \mathrm{I}$, the identity matrix and $\Upsilon_1 = \Phi_1\Phi_2$ are linearly independent. If not, then
\begin{eqnarray*}
\mathrm{I} &=& \alpha\Phi_1\Phi_2, \,\,\,\alpha \neq 0,\\
\implies S_2 &\backsimeq& \alpha S_2\Phi_1\Phi_2\\
&\backsimeq& S_2\Phi_2,
\end{eqnarray*}
a contradiction by (\ref{subspaceconditions2}).

\emph{Inductive step:} Suppose that the claim is true for $s$ and consider the case $s+1$.
If the claim is false for $s+1$, then some linear combination of elements in $M_{s+1}$ is equal to zero. Note that each element in $M_{s+1}$ is a product ending either in $\Phi_{2s+1}\Phi_{2s+2}$ or not. We have:
\begin{eqnarray}\label{induc}
\Psi_1^{(s)} = \Psi_2^{(s)}\Phi_{2s+1}\Phi_{2s+2},
\end{eqnarray}
where $\Psi_1^{(s)}$ and $\Psi_2^{(s)}$ are linear combinations of elements in $M_s$. Now, operating both sides on the subspace $S_{2s+2}$, we obtain
\begin{eqnarray}
S_{2s+2}\Psi_1^{(s)} &\backsimeq& S_{2s+2}\Psi_2^{(s)}\Phi_{2s+1}\Phi_{2s+2}.
\end{eqnarray}
As in Theorem \ref{thm-lincomb2subsp}, this is possible only if 
\begin{eqnarray}\label{induc2}
S_{2s+2}\Psi_1^{(s)} &=& \{0\}.
\end{eqnarray} 
Similarly, operating both sides of (\ref{induc}) on the subspace $S_{2s+1}$, we obtain 
\begin{eqnarray}\label{induc3}
S_{2s+1}\Psi_1^{(s)} &\backsimeq& S_{2s+1}\Psi_2^{(s)}\Phi_{2s+1}\Phi_{2s+2}.
\end{eqnarray}
The left hand side of the equation is in $S_{2s+1}$ and the right hand side is in $S_{2s+1}\Phi_{2s+1}\Phi_{2s+2}$. But,
\begin{eqnarray}
S_{2s+1} \oplus S_{2s+1}\Phi_{2s+1} &\backsimeq& \mathbb{F}^{\ell},\,\,(\mathrm{cf.}\,\,(\ref{subspaceconditions2}))\nonumber\\
\implies S_{2s+1}\Phi_{2s+2} \oplus S_{2s+1}\Phi_{2s+1}\Phi_{2s+2}&\backsimeq& \mathbb{F}^{\ell},\nonumber\\
\implies S_{2s+1} \oplus S_{2s+1}\Phi_{2s+1}\Phi_{2s+2}&\backsimeq& \mathbb{F}^{\ell}.\,\,(\mathrm{cf.}\,\,(\ref{subspaceconditions1}))\nonumber
\end{eqnarray} 
Thus, (\ref{induc3}) is possible only if
\begin{eqnarray}\label{induc4}
S_{2s+1}\Psi_1^{(s)} &=& \{0\}.
\end{eqnarray}
But by our assumption, $S_{2s+1} \cap S_{2s+2} = \{0\}$. So, by (\ref{induc2}), (\ref{induc4}), we have $\Psi_1^{(s)} = 0$, contradicting our induction assumption.
\end{IEEEproof}

\smallskip
We are now ready to prove the main theorem of this section.

\begin{thm}\label{thm2}
For a given $\ell$ and $k$, if there exist invertible matrices $\Phi_i, i \in [k]$ and subspaces $S_i, i \in [k]$, satisfying the subspace conditions (\ref{subspaceconditions1}) and (\ref{subspaceconditions2}), then
\begin{eqnarray}
k \le \max\left(4\ell,8\log_2\ell\right).
\end{eqnarray}
\end{thm}
\begin{IEEEproof}
We are given a set of subspaces $S_i, i \in [k]$ and the corresponding set of matrices $\Phi_i, i \in [k]$, which satisfy the subspace conditions (\ref{subspaceconditions1}) and (\ref{subspaceconditions2}). Suppose we can find at most $n$ disjoint pairs of complementary subspaces $(S_i,S_j)$ in the given set and no more, where $2n \le k$. Without loss of generality, let these pairs be $(S_1,S_2), \ldots, (S_{2n-1},S_{2n})$. Let $k$ be an even integer for convenience. We can construct a set $T$ as in Theorem \ref{thm-lincomb2subsp}, where $T$ is the set of $((k-2n)/2)^2$ products of the form $\Phi_{2p-1}\Phi_{2q}$, where $2p-1,2q \in \{2n+1,\ldots,k\}$. Note that we could have arranged the $(k-2n)$ matrices $\Phi_i, i \in \{2n+1, \ldots, k\}$ into any two subsets of size $(k-2n)/2$ and taken products where the first factor is in the first set and the second factor in the second set.

Observe that by Corollary \ref{cor1}, the set $T$ must consist of $((k-2n)/2)^2$ linearly independent matrices. Otherwise, we can find another pair of complementary subspaces disjoint from the given $n$ pairs. Let $R$ be the following set of $2^n\,((k-2n)/2)^2$ $\ell \times \ell$ matrices:
\begin{eqnarray*}
\Upsilon_{\epsilon_1\epsilon_2\ldots\epsilon_n}^{i}=\Omega_{i}\prod_{j=1}^{n}\left(\Phi_{2j-1}\Phi_{2j}\right)^{\epsilon_j},
\end{eqnarray*}
where $\epsilon_j \in \{0,1\}$ for all $j \in [n]$, $i \in [|T|] := \{1,\ldots,|T|\}$. $\Omega_i$ is the $i^{\mathrm{th}}$ matrix in the set $T$.

It can be proved that $R$ consists of $|R|$ linearly independent $\ell \times \ell$ matrices. The proof runs along the same lines as in Theorem \ref{thm-completingsubsp2}, except that the base case relies on the linear independence of the matrices in $T$. Notice that $S_j\Omega_i \backsimeq S_j$, where $S_j, j \in [2n]$, is a subspace in the list of complementary subspaces and $\Omega_i, i \in \{2n+1,\ldots,k\}$, is a matrix in the set $T$.

We therefore have\footnote{Note that if $n = k/2$, we have $|R| = 2^{k/2}$ and $T$ is an empty set.} $|R| = 2^n\,((k-2n)/2)^2$ linearly independent $\ell \times \ell$ matrices, and as in Theorem \ref{thm1}, to satisfy dimensionality,
\begin{eqnarray}
|R| \le \ell^2.
\end{eqnarray}

The only missing link is that we do not really know how many complementary subspaces we can find. A simple bound can be obtained by taking the cases $n \le k/4$ and $n > k/4$, one of which must necessarily occur. If $n \le k/4$ and for convenience, say $k$ is a multiple of $4$ and $k \ge 8$, then $|R| \ge k^2/16$. If $n > k/4$, then $|R| \ge 2^{k/4}$. Thus,
\begin{eqnarray*}
\min\left(2^{k/4}, k^2/16\right) &\le& \ell^2,\,\,\mathrm{or},\\
k &\le& \max\left(4\ell,8\log_2\ell\right).
\end{eqnarray*}
It can be shown that for $\ell > 7$, we have $k \le 4\ell$.

A tighter bound can be obtained by observing that
\begin{eqnarray*}
\min\left(\min_{n \in \{0,1,\ldots,k/2-1\}} 2^n\,((k-2n)/2)^2, 2^{k/2}\right) \le \ell^2,
\end{eqnarray*}
which for a sufficiently large $k$, results in the bound $k \le 2\ell$.
\end{IEEEproof}

\section{Closing In: A Log-Squared Bound}\label{sec:ub3}
We harnessed the intersection properties of two subspaces in Section \ref{sec:ub2} to obtain an upper bound on $k$ which is linear in $\ell$. Specifically, we used the fact that the trivial intersection of two subspaces is equivalent to their sum spanning the entire space. This ceases to hold for more than two subspaces. For instance, it is not true that if three subspaces of dimension $\ell/2$ intersect trivially, then their sum spans the entire space. As has been mentioned before, the ``trivial intersection'' thread of analysis can be generalized as in Appendix \ref{appendix:cor1}.
In this section, we extend the ``spanning of the whole space'' thread.
Despite the lack of an either-or condition for the two properties which we used in Theorem \ref{thm2} in the previous section, we show at the end of this section that for two parity nodes, we can upper bound $k$ by $2\left(\log_2\ell\right)\left(\log_2\ell+1\right)$. Unless mentioned otherwise, we consider only two parity nodes ($r =2$) in this section and make use of the subspace conditions (\ref{subspaceconditions1}) and (\ref{subspaceconditions2}).

\smallskip
We now explore the consequences of finding subspaces whose sum is the full vector space $\mathbb{F}^{\ell}$.
\begin{thm}[Extension of Theorem \ref{thm-completingsubsp2}]\label{thm-completingsubsp-r}
Let ${\cal X}_1, \ldots, {\cal X}_n$ be a partition of the set of integers $[k]$ into $n$ sets, such that for any set ${\cal X}_i$ in the partition, we have 
%
\begin{eqnarray}
\label{assum2}
\sum_{j\in \cX_i}S_j &=&\mathbb{F}^{\ell}.
\end{eqnarray}
Let $t_i=\sum_{j=1}^{i}|\cX_j|$ be the sum of sizes of the first $i$ sets $\cX_i$, and without loss of generality assume that 
$\cX_i=\{t_{i-1}+1,\ldots,t_i\}.$
For $i\in [n]$, define 
$$\Lambda_i := \prod_{j=t_{i-1}+1}^{t_i}\Phi_{j},$$
as the product in an ascending order of the matrices with indices in $\cX_i$.
Then the following set of $2^n$ square matrices of order $\ell$, 
\begin{eqnarray*}
\Upsilon_{\epsilon_1\epsilon_2\ldots\epsilon_n}=\prod_{i=1}^{n}\Lambda_i^{\epsilon_i},\,\,\textrm{where}\,\,\epsilon_i \in \{0,1\}, \textrm{ for all } i \in [n], \,\,
\end{eqnarray*}
are linearly independent.
\end{thm}
\begin{IEEEproof}
The proof proceeds as in Theorem \ref{thm-completingsubsp2} using induction. 
 
\smallskip
\emph{Induction Claim:} For all $s \in [n]$. the following set of $2^s$ square matrices of order $\ell$, 
\begin{eqnarray*}
\Upsilon_{\epsilon_1\epsilon_2\ldots\epsilon_s}=\prod_{p=1}^{s}\Lambda_p^{\epsilon_p},\,\,\,\textrm{where}\,\,\epsilon_p \in \{0,1\}, \textrm{ for all } p \in [s], \,\,
\end{eqnarray*}
are linearly independent.

\emph{Base case:} The identity matrix $\Upsilon_0 = \mathrm{I}$ and $\Upsilon_1 = \Phi_1\Phi_2\cdots\Phi_{t_1} = \Lambda_1$ are linearly independent because while $S_{t_1} \mathrm{I} \backsimeq S_{t_1}$, we have $S_{t_1} \Lambda_1 \backsimeq S_{t_1}\Phi_{t_1}$, which does not intersect with $S_{t_1}$.

\emph{Inductive step:} Let the inductive claim hold for some $s$, then it is true for $s+1$. Otherwise, we have
\begin{eqnarray}\label{induc21}
\Psi_1^{(s)} &=& \Psi_2^{(s)}\Lambda_{s+1},\,\,\,\textrm{or},\\
\Psi_1^{(s)} &=& \Psi_2^{(s)}\Phi_{t_s+1}\Phi_{t_s+2}\cdots\Phi_{t{s+1}},
\end{eqnarray}
where $\Psi_1^{(s)}$ and $\Psi_2^{(s)}$ are linear combinations of elements of the form $\Upsilon_{\epsilon_1\epsilon_2\ldots\epsilon_s}$. Operating both sides on the subspace $S_{i}$ for some $i$ in  $\cX_{s+1}$,
\begin{eqnarray}\label{aux11}
S_{i}\Psi_1^{(s)} &\backsimeq& S_{i}\Psi_2^{(s)}\Phi_{t_s+1}\Phi_{t_s+2}\cdots\Phi_{t_{s+1}}.
\end{eqnarray}
The right hand side in (\ref{aux11}) lies in $S_{i}\Phi_{i}\Phi_{i+1}\cdots\Phi_{t_{s+1}}$, whereas the left hand side lies in $S_{i} = S_{i}\Phi_{i+1}\cdots\Phi_{t_{s+1}}$, which is a complementary subspace to $S_{i}\Phi_{i}\Phi_{i+1}\cdots\Phi_{t_{s+1}}$. Thus, we have, 
\begin{eqnarray*}
S_{i}\Psi_1^{(s)} &=& \{0\}, \,\,\,\forall\,i\in {\cal X}_{s+1}, \,\,\textrm{or},\\
\left(S_{t_s+1} + \cdots + S_{t_{s+1}}\right)\Psi_1^{(s)} &=& \{0\},
\end{eqnarray*}
which, from (\ref{assum2}), implies that $\Psi_1^{(s)}=0$, contradicting the induction assumption.
\end{IEEEproof}

\smallskip
\textcolor{black}{In order to apply the previous theorem, we would like to find subspaces whose sum is $\mathbb{F}^{\ell}$. The following theorem and its corollary 
show that in fact \emph{any} sum of a small number of subspaces equals $\mathbb{F}^{\ell}$}.
\begin{thm}\label{thm:sum}
Given the subspace conditions (\ref{subspaceconditions1}) and (\ref{subspaceconditions2}), consider any set of $n$ subspaces $S_{i_1}, S_{i_2}, \ldots, S_{i_n}$, where $i_j \in [k]$.
The dimension of the sum of these subspaces satisfies
\begin{eqnarray}
\dim\left(S_{i_1} + \ldots + S_{i_n}\right) &\ge& \left(1-\frac{1}{2^n}\right)\ell.
\end{eqnarray}
\end{thm}

\begin{IEEEproof}
Again, we turn to induction.

\emph{Base case:} The dimension of $S_{i_1}$ is $\ell/2$.

\emph{Inductive step:} Suppose the theorem is true for $s$. We assume without loss of generality that $i_j = j$. 
Let $S=S_1+\cdots+S_s$. Note that  
\begin{eqnarray}
\dim(S\cap S_{s+1}) &\leq& \frac{\dim(S)}{2}.
\label{eq:good}
\end{eqnarray}
If we assume the contrary, since $S$ is an invariant subspace of $\Phi_{s+1}$, we have $(S \cap S_{s+1})\Phi_{s+1}\subseteq S$.
On the other hand $S\cap S_{s+1}$ is also a subspace of $S_{s+1}$, therefore
\begin{eqnarray*}
\begin{array}{ccccc}
\{0\} &\neq& (S\cap S_{s+1})\cap (S\cap S_{s+1})\Phi_{s+1} &\subseteq& S_{s+1}\cap S_{s+1}\Phi_{s+1},
\end{array}
\end{eqnarray*}
which contradicts 
\eqref{subspaceconditions2}.
We conclude then that 
\begin{eqnarray*}
\dim(S+S_{s+1})&=&\dim(S)+\dim(S_{s+1})-\dim(S\cap S_{s+1}),\\[0.2cm]
&\geq& \frac{\dim(S)}{2}+ \frac{\ell}{2},\\[0.1cm]
&\geq& \left(1-\frac{1}{2^{s+1}}\right)\ell,
\end{eqnarray*}
where the final inequality uses the induction assumption.
\end{IEEEproof}

\smallskip
\begin{cor}\label{cor:sum}
For $n = \log_2\ell + 1$,
\begin{eqnarray}
\dim\left(S_{i_1} + \cdots + S_{i_n}\right) &=& \mathbb{F}^{\ell}.
\end{eqnarray}
\end{cor}
\begin{IEEEproof}
From Theorem \ref{thm:sum}, we have for $n = \log_2\ell+1$,
\begin{eqnarray*}
2\dim\left(S_{i_1} + \cdots + S_{i_n}\right) &\ge& (2\ell-1), \,\,\,\textrm{or},\\
\dim\left(S_{i_1} + \cdots + S_{i_n}\right) &=& \ell.
\end{eqnarray*}
\end{IEEEproof}

\smallskip
We finally apply Theorem \ref{thm-completingsubsp-r} to obtain the tighter bound in Theorem \ref{thmlogsq}.
\smallskip
\begin{thm}\label{thmlogsq}
For any given optimal bandwidth $(k+3, k+1, \ell)$ MDS code, with $r = 2$ parity disks, the following upper bound holds:
\begin{eqnarray}
\begin{array}{ccccc}
k &\le& 2\left(\log_2\ell\right)\left(\log_2\ell+1\right) &=& (2+o(1))\log_2^2\ell.
\end{array}
\end{eqnarray}
\end{thm}
\begin{IEEEproof}
From Theorem \ref{thm-constant-repairing-subspaces}, we know that if there exists an optimal bandwidth $(k+3, k+1, \ell)$ MDS code, then there exists a set of invertible matrices $\Phi_1, \ldots, \Phi_k$ of order $\ell$ and a corresponding set of subspaces $S_1, \ldots, S_k$, each of dimension $\ell/2$ such that the sets satisfy the subspace conditions (\ref{subspaceconditions1}) and (\ref{subspaceconditions2}).

Partition the set $[k]$ into sets of size $t = \log_2\ell + 1$ (we assume that $t$ divides $k$). From Corollary \ref{cor:sum}, each set in the partition satisfies the conditions in Theorem \ref{thm-completingsubsp-r}. We therefore have $2^{k/t}$ linearly independent $\ell \times \ell$ matrices, and as in Theorem \ref{thm1}, to satisfy dimensionality, we have
%
%
\begin{eqnarray*}
2^{k/(\log_2\ell+1)} &\le& \ell^2,\,\,\,\textrm{or},\\
k &\le& 2\left(\log_2\ell\right)\left(\log_2\ell+1\right).
\end{eqnarray*}
\end{IEEEproof}

The proof of a log-squared bound in $\ell$ on the number of systematic nodes $k$ for an arbitrary number of parity nodes $r$ follows an identical line of reasoning as for two parity nodes. This and the corresponding extension of the theorems in this section for any $r$ are given in Appendix \ref{appendix:all-parity}.

\section{Conclusion}\label{sec:conc}
In this paper, we make critical progress towards the open problem of finding the maximum number of systematic storage disks $k$ for which an optimal bandwidth MDS code exists, for a given sub-packetization $\ell$ and number of parity nodes $r$. We show that this $k$ is bounded from above by a log-squared function of $\ell$, thereby almost closing the existing optimality gap in \cite{TWB13}. A practical consequence of this result is that the maximum $k$ does not improve significantly than for the case of MDS codes with optimal access. In the latter case, the parameter of interest is the number of symbols {\em read} during the repair process rather than the number of symbols {\em transmitted} to the repair center, which is typically smaller. The maximum $k$ here is known \cite{TWB13} to be $r\log_r\ell$. 

Of more general interest, we introduce previously unexploited methods of analysis by translating the storage problem into a geometric problem involving a set of operators and subspaces. We posit that such a geometric analysis involving ideas of linear independence may be useful for attacking other open problems in distributed storage and elsewhere. For example, the maximum file size which can be securely stored on an optimal bandwidth MDS code has been derived in \cite{GRCP13}. The eavesdropper is assumed to be passive (that is, does not modify the eavesdropped symbols), and to have access to any given number of systematic nodes and the vectors transmitted during their corresponding repair processes. The underlying analysis uses Theorem \ref{thm:sum} and closes the optimality gap for this case and proves that a zigzag code precoded by a maximum rank distance code achieves the maximum file size \cite{RKSV12, TWB13ZZ}.

{\em Open Problems}: The problem of finding the maximum possible $k$ that we set out to achieve remains open. We think that it may be possible to improve the upper bound in Theorem \ref{thmlogsq-r} for the case of arbitrary number of parity nodes $r$ using results which utilize {\em all} the encoding matrices (and not just a set corresponding to one parity node).

\bibliographystyle{IEEEbib_wrc}
\bibliography{IEEEabrv,references}

\appendices
\section{Generalization of Corollary \ref{cor1}}\label{appendix:cor1}

\begin{thm}
\label{thm1ext}
Let ${\cal O}_{1}, \ldots, {\cal O}_{t}$ be an equally sized partition of the set of integers $[k]$. Let $\Gamma$ be the set of all vectors $(i_1,\ldots,i_t)$ in ${\cal O}_{1}\times \cdots\times {\cal O}_{t}$ such  that the corresponding subspaces do not intersect trivially, namely
\begin{eqnarray*}
\Gamma &=& \left\{\left(i_1,\ldots,i_t\right) \in {\cal O}_1\times \cdots\times {\cal O}_t \left| \,S_{i_1} \cap \ldots \cap S_{i_t}\neq\{0\}\right\}.\right. 
\end{eqnarray*}
Then the $|\Gamma|$ matrices defined as 
$$\prod_{j=1}^t\Phi_{i_j}, \textrm{ such that }\left(i_1, \ldots, i_t\right) \in \Gamma,$$
are linearly independent. 
\end{thm}
\begin{IEEEproof}
For a vector $v$ in $\Gamma$, denote by $v_i$ its $i^{\mathrm{th}}$ coordinate. Suppose the claim is not true and we have without loss of generality
\begin{eqnarray*}
\prod_{i=1}^t\Phi_i &=& -\sum_{\substack{v \in \Gamma\\ v_t=t}}\alpha_v\left(\prod_{j=1}^{t-1}\Phi_{v_j}\right)\Phi_t + \sum_{\substack{v \in \Gamma\\ v_t\neq t}}\alpha_{v}\prod_{j=1}^{t}\Phi_{i_j}.
\end{eqnarray*}
Following the flavor of argument in Theorem \ref{thm-lincomb2subsp}, we can partition the terms in the above summation into those for which $S_t$ is invariant and those which take $S_t$ to the complementary subspace $S_t\Phi_t$. Taking the latter set of terms, we have
\begin{eqnarray*}
S_t\left(\prod_{i=1}^{t-1}\Phi_i + \sum_{\substack{v \in \Gamma\\ v_t=t}}\alpha_{v}\prod_{j=1}^{t-1}\Phi_{i_j}\right)\Phi_t &=& \{0\}.
\end{eqnarray*}
Applying the non-singularity of $\Phi_t$, we have
\begin{eqnarray*}
S_t\left(\prod_{i=1}^{t-1}\Phi_i + \sum_{\substack{v \in \Gamma\\ v_t=t}}\alpha_{v}\prod_{j=1}^{t-1}\Phi_{i_j}\right) &=& \{0\}, \,\,\,\textrm{and thus}\\
\left(S_{t-1}\cap S_t\right)\left(\prod_{i=1}^{t-1}\Phi_i + \sum_{\substack{v \in \Gamma\\ v_t=t}}\alpha_{v}\prod_{j=1}^{t-1}\Phi_{i_j}\right) &=& \{0\}.
\end{eqnarray*}

Again, we can partition the terms (operators) within the second set of parentheses into those for which $S_{t-1} \cap S_t$ is an invariant subspace and those which take it to a non-intersecting subspace $S_{t-1}\Phi_{t-1}\cap S_t$. As before, we take the latter set of terms to obtain
\begin{eqnarray*}
\left(S_{t-1}\cap S_t\right)\left(\prod_{i=1}^{t-2}\Phi_i + \sum_{\substack{v \in \Gamma\\ v_{t-1}=t-1\\v_t=t}}\alpha_{v}\prod_{j=1}^{t-2}\Phi_{i_j}\right)\Phi_{t-1} &=& \{0\},
\end{eqnarray*}
and by non-singularity of $\Phi_{t-1}$,
\begin{eqnarray*}
\left(S_{t-1}\cap S_t\right)\left(\prod_{i=1}^{t-2}\Phi_i + \sum_{\substack{v \in \Gamma\\ v_{t-1}=t-1\\v_t=t}}\alpha_{v}\prod_{j=1}^{t-2}\Phi_{i_j}\right) &=& \{0\}.
\end{eqnarray*}

Iterating this argument, we finally obtain
\begin{eqnarray*}
\left(S_1 \cap \ldots \cap S_t\right)\Phi_1 = \{0\},
\end{eqnarray*}
or, by the non-singularity of $\Phi_1$,
\begin{eqnarray*}
S_1 \cap \ldots \cap S_t = \{0\},
\end{eqnarray*}
a contradiction per the definition of $\Gamma$.
\end{IEEEproof}

\section{Log-Squared Bound for $r$ Parity Nodes}\label{appendix:all-parity}
As in the case of two parity nodes, we first derive a lower bound on the dimension of the sum of repairing subspaces for the transformed $(k+r, k,\ell)$ MDS code, for which the repairing subspaces are independent of the helper node.
\begin{thm}[Extension of Theorem \ref{thm:sum}]
Given the subspace conditions (\ref{sptx1}) and (\ref{sptx2}), consider any set of $n$ subspaces $S_{i_1}, S_{i_2}, \ldots, S_{i_n}$, of dimension $\ell/r$, where $i_j \in [k]$, for all $j \in [n]$. The dimension of the sum of these subspaces satisfies
\begin{eqnarray}
\dim\left(S_{i_1} + \ldots + S_{i_n}\right) &\ge& \left(1-\left(\frac{r-1}{r}\right)^n\right)\ell.
\end{eqnarray}
\end{thm}
\begin{IEEEproof} We apply induction on $n$.
Without loss of generality, we can assume $i_j = j$, for all $j \in [n]$.

\emph{Base case:} The theorem is true for $n = 1$ because
$\dim\left(S_{1}\right) = \ell/r$.

\emph{Inductive step:} Suppose that the theorem is true for $n$. Denoting the subspace $S_1 + \cdots + S_n$ by $S$, we can prove that
\begin{eqnarray}
\label{sub-int-aux}
\dim\left(S \cap S_{n+1}\right) &\le& \frac{\dim(S)}{r}.
\end{eqnarray}
Notice that
\begin{eqnarray*}
\dim\left(\left(S \cap S_{n+1}\right)A_{t,n+1}\right) &=& \dim\left(S \cap S_{n+1}\right),
\end{eqnarray*}
for all $t \in [r]$, and that the subspace, 
\begin{eqnarray*}
\left(S \cap S_{n+1}\right)A_{t,n+1} &\backsimeq& S \cap S_{n+1}A_{t,n+1},
\end{eqnarray*}
is a subspace of both $S$ and $S_{n+1}A_{t,n+1}$. By the subspace condition (\ref{sptx2}), we know that the subspaces $S_{n+1}A_{1,n+1}, \ldots, S_{n+1}A_{r,n+1}$ are mutually trivially intersecting, and therefore
\begin{eqnarray*}
\sum_{t=1}^{r}\dim\left(\left(S \cap S_{n+1}\right)A_{t,n+1}\right) &=& r\dim\left(S \cap S_{n+1}\right),\\
&\le& \dim\left(S\right),
\end{eqnarray*}
thereby proving  (\ref{sub-int-aux}). Finally we have,
\begin{eqnarray*}
\dim\left(S + S_{n+1}\right) &=& \dim\left(S\right) + \dim\left(S_{n+1}\right) - \dim\left(S \cap S_{n+1}\right),\\[0.2cm]
&\ge& \frac{r-1}{r}\dim\left(S\right) + \dim\left(S_{n+1}\right),\\[0.1cm]
&\ge& \left(1-\left(\frac{r-1}{r}\right)^{n+1}\right)\ell,
\end{eqnarray*}
where the last inequality uses the induction assumption.
\end{IEEEproof}

\begin{cor}\label{cor:sum-r}[Extension of Corollary \ref{cor:sum}]
For $n = \lfloor\log_{\delta}\ell\rfloor + 1$, where $\delta = r/(r-1)$,
\begin{eqnarray}
\dim\left(S_{i_1} + \cdots + S_{i_n}\right) &=& \mathbb{F}^{\ell}.
\end{eqnarray}
\end{cor}

\smallskip
Finally, applying Theorem \ref{thm-completingsubsp-r}, we  obtain the log-squared bound in Theorem \ref{thmlogsq-r}.
\smallskip
\begin{thm}\label{thmlogsq-r}[Extension of Theorem \ref{thmlogsq}]
For any given optimal bandwidth $(k+r+1,k+1,\ell)$ MDS code, with $r$ parity disks, the following upper bound holds:
\begin{eqnarray*}
\begin{array}{ccccc}
k &\le& 2\left(\log_2\ell\right)\left(\lfloor\log_{\delta}\ell\rfloor+1\right) &=& O(\log^2\ell),
\end{array}
\end{eqnarray*}
where $\delta = r/(r-1)$. 
\end{thm}
\begin{IEEEproof}
The proof is exactly the same as for Theorem \ref{thmlogsq}. As mentioned in Section \ref{sec:problem}, we can assume $A_{2,j}$ as an identity matrix for all $j \in [k]$. Let us define $A_{1,j} = \Phi_j$ for all $j \in [k]$. This relabeling leads to the following subspace conditions:
\begin{eqnarray}
\label{subspaceconditions-one}
S_i\Phi_j &\backsimeq& S_i, \mathrm{\,\,\,and}\\
\label{subspaceconditions-two}
S_i\Phi_i \cap S_i &\backsimeq& \{0\}.
\end{eqnarray}
Notice that they are simply a subset (relaxation) of the conditions (\ref{sptx1}) and (\ref{sptx2}).
To complete the proof, we can now use Theorem \ref{thm-completingsubsp-r} which incidentally only uses (\ref{subspaceconditions-two}) instead of (\ref{subspaceconditions2}).
\end{IEEEproof}

\end{document}